# Validation of Monte Carlo $^{131}$I radiopharmaceutical dosimetry workflow using a 3D printed anthropomorphic head and neck phantom

David P. Adam[1], Joseph Grudzinski[2], Ian Bormett[3], Benjamin L. Cox[3], Ian R. Marsh[1], Tyler J. Bradshaw[2], Paul Harari[4], Bryan Bednarz[1a]

[1)] Department of Medical Physics, University of Wisconsin-Madison, Madison, WI, 53705
[2)] Department of Radiology, University of Wisconsin-Madison, Madison, WI, 53705
[3)] Morgridge Institute for Research, University of Wisconsin-Madison, Madison, WI, 53705
[4)] Department of Human Oncology, University of Wisconsin-Madison, Madison, WI, 53705
[a)] Author to whom correspondence should be addressed.

To Medical Physics

**Grant support:** This project was partially supported by the Morgridge Institute for Research, the Specialized Program of Research Excellence (SPORE) program, through the NIH National Institute for Dental and Craniofacial Research (NIDCR) and National Cancer Institute (NCI), grant P50DE026787, the National Cancer Institute (NCI) Research Project--Cooperative Agreements U01CA233102-01, and the NIH National Cancer Institute (NCI), grant P01CA250972.

**Corresponding author:** Bryan P. Bednarz, Department of Medical Physics, 1005 WIMR, 1111 Highland Avenue, Madison WI 53705; +1-608-262-5225; bbednarz2@wisc.edu

**First author:** David P. Adam (graduate student), Department of Medical Physics, 1005 WIMR, 1111 Highland Avenue, Madison WI 53705; +1-(608) 265-6116; dadam@wisc.edu

**Word Count:** 5411 (excluding references)

**Running title:** $^{131}$I anthropomorphic phantom dosimetry

**Abstract**

**Purpose:**

Approximately 50% of head and neck cancer (HNC) patients will experience loco-regional disease recurrence following initial courses of therapy. Retreatment with external beam radiotherapy (EBRT) is technically challenging and accompanied by a significant risk of irreversible damage to normal tissues. Radiopharmaceutical therapy (RPT) is a potential method to treat recurrent HNC in conjunction with EBRT. Phantoms are used to calibrate and add quantification to nuclear medicine images and anthropomorphic phantoms can account for both the geometrical and material composition of the head and neck. In this study, we present the creation of an anthropomorphic, head and neck, nuclear medicine phantom and its characterization for the validation of a Monte Carlo, SPECT image based, $^{131}$I RPT dosimetry workflow.

**Methods:**

3D printing techniques were used to create the anthropomorphic phantom from a patient CT dataset. Three $^{131}$I SPECT/CT imaging studies were performed using a homogeneous, Jaszczak, and an anthropomorphic phantom to quantify the SPECT images using a GE Optima NM/CT 640 with a high energy general purpose (HEGP) collimator. The impact of collimator detector response (CDR) modeling and volume-based partial volume corrections (PVC) upon the absorbed dose was calculated using an image based, Geant4 Monte Carlo RPT dosimetry workflow and compared against a ground truth scenario. Finally, uncertainties were quantified in accordance with recent EANM guidelines.

**Results:**

The 3D printed anthropomorphic phantom was an accurate re-creation of patient anatomy including bone. The extrapolated Jaszczak recovery coefficients were greater than that of the 3D printed insert (~22.8 ml) for both the CDR and non-CDR cases (with CDR: 0.536 vs. 0.493, non-CDR: 0.445 vs. 0.426 respectively). Utilizing Jaszczak phantom PVCs, the absorbed dose was underpredicted by 0.7% and 4.9% without and with CDR, respectively. Utilizing anthropomorphic phantom RCs overpredicted the absorbed dose by 3% both with and without CDR. All dosimetry scenarios that incorporated PVC were within the calculated uncertainty of the activity. The uncertainties in the cumulative activity ranged from 25.6% to 113% for Jaszczak spheres ranging in volume from 0.5 ml to 16 ml.

**Conclusion:**

The accuracy of Monte Carlo-based dosimetry for $^{131}$I RPT in head and neck cancer was validated with an anthropomorphic phantom. In this study, it was found that Jaszczak based PVC were sufficient. Future applications of the phantom could involve 3D printing and characterizing patient specific volumes for more personalized RPT dosimetry estimates.

# 1. Introduction

Approximately 50% of head and neck cancer (HNC) patients will experience loco-regional disease recurrence following initial courses of therapy[1–6]. Although a majority of patients can be cured from HNC, the 5 year overall survival rate of patients with locally advanced and/or recurrent HNC is approximately 50-60%[7] and cure rates have improved only marginally over the last 30 years[8]. Retreatment of HNC is technically challenging and accompanied by a significant risk of irreversible damage to normal tissues that can translate into profound adverse effects on patient health-related quality of life[9,10,19–21,11–18]. Several studies have ascertained that, despite improvement in long term survival (2 years) in approximately 10-20% of recurrent HNC patients following external beam radiation therapy (EBRT), there is considerable risk for high-grade toxicities after treatment [9,10,19–22,11–18]. Despite the ability of EBRT to provide disease control, efforts to reduce the potential of acute and long-term tissue injury in locoregionally recurrent HNC patients following EBRT are clearly warranted.

Radiopharmaceutical therapy (RPT) is one potential treatment that can be combined with EBRT to maintain disease control of recurrent HNC and mitigate high-grade toxicities. RPT involves a biological delivery vector labeled with a radioactive isotope that is designed to preferentially attach to or be internalized by cancer cells. The radioactive drug conjugate is infused directly into a patient's bloodstream allowing it to reach tumor sites located throughout the body. Due to this unique delivery route, the dose limiting organs are different than EBRT; and, when combined with EBRT, the therapeutic window in the patient can be expanded, allowing physicians to treat more aggressively.

CLR 131 [23]. an alkyl phosphocholine (APC) analog with broad cancer targeting abilities, is a promising RPT agent paired to the radioisotope $^{131}$I, which emits both therapeutic beta particles (mean energy of 606 keV) and diagnostic gamma rays (364 keV) which can be localized using SPECT. APCs enter cells through specialized plasma membrane microdomains called lipid rafts [23], which are expressed 6-10 times more in cancer cells in comparison to normal tissues. The increased specificity of CLR 131 for malignant cells has led to increased therapeutic efficacy in both preclinical and clinical studies. Preclinical studies confirm preferential uptake and retention of CLR 131 in murine HNC xenografts and tumor growth inhibition[24]. Tumor targeting has been observed in patients in on-going and completed clinical trials [25–27]. Additional tumor growth inhibition has been demonstrated when CLR 131 is combined with EBRT[24].

Given these results, a clinical trial (NCT04105543) is underway at UW-Madison to investigate the efficacy of combining CLR 131 with EBRT with curative intent in treating relapsed HNC [28].

The premise of the trial is to evaluate the combination of CLR 131 and EBRT in the treatment of recurrent HNC wherein the radiation dose deposited by the CLR 131 permits a dose reduction delivered by EBRT (e.g., standard of care 60-70 Gy will be targeted irrespective of radiation modality). The overall hypothesis of the clinical trial is that the addition of CLR 131 will support a dose reduction of EBRT which will is safe and tolerable while maintaining favorable tumor response rates and diminishing the adverse impact of radiation treatment on subject specific symptoms, such as quality of life, salivary flow and swallowing function.

To achieve the objectives of the trial [29], patient-specific dosimetry is needed to accurately characterize the tumor absorbed dose from CLR 131 on the voxel level. Longitudinal SPECT/CT [31] scans are used to map the time-dependent biodistribution of CLR 131 in the patient and serve as the input for voxel-level dosimetry calculations performed by an in-house Monte Carlo-based software called RAPID [30]. The accuracy of the CLR 131 dose calculation depends on the quantitative accuracy of $^{131}$I SPECT images. Nuclear medicine phantom scans are used both for quantitative calibration (i.e. convert counts to units of activity) and correction of partial volume effects (PVE). PVEs, which are dependent on both the size and shape of the structure being imaged [31–33], are caused by inherent resolution limitations of imaging systems and can significantly degrade SPECT/CT image quality and quantitative accuracy.

Collimator detector response (CDR) modeling has been recommended by MIRD pamphlet 24 [34] for SPECT image reconstruction due to septal penetration of high energy $^{131}$I photons through the collimator. Without CDR modeling, noise can be quite apparent on reconstructed images and small structures can be difficult to resolve; however, some have observed Gibbs ringing artifacts on images reconstructed with CDR enabled which can affect the resultant activity distribution [35].

Anthropomorphic nuclear medicine phantoms have been investigated for a variety of organs (kidney [36–39], liver [37,39–41], spleen [37,39], and pancreas[37,38]), but few groups have investigated the use of nuclear medicine head and neck phantoms. 3D printed head and neck phantoms have been created for CT and MR applications but few for nuclear medicine applications. A prototype head and neck phantom was described by Alqahtani et. al in which the performance of gamma imaging systems was evaluated but the study did not extend to dosimetry [42]. In these studies, results have generally indicated that anthropomorphic phantoms account for heterogeneities that are not present in traditional nuclear medicine phantoms and are often more accurate than non-anthropomorphic phantoms. 3D printed anthropomorphic head and neck phantoms that account for the changing contour of the head and neck region and the heterogeneous material composition in the patient have not been investigated. A couple of recently published review articles succinctly demonstrate the state of 3D printed phantoms [43,44].

In this work we aim to validate our image based RPT dosimetry workflow for the HNC clinical trial with the introduction and characterization of a novel, 3D printed, anthropomorphic, head and neck nuclear medicine phantom that mimics the geometry and material composition of the head and neck region. Imaging studies will be conducted with both traditional Jaszcazak nuclear medicine phantoms and the anthropomorphic phantom to investigate $^{131}$I SPECT/CT calibration, partial volume corrections (PVCs), and the role of collimator detector response (CDR) modeling on dosimetry. The resultant data from the imaging studies will then be used as input for the $^{131}$I RPT dosimetry workflow and compared to ground truth scenarios. Finally, the uncertainties from the imaging studies will be investigated and the impact of uncertainties on the dosimetry will be discussed.

## 2. Methods
### 2.1 Phantom Creation

The original CT data for the anthropomorphic phantom was accessed from a publicly available database provided by Radiation Therapy Oncology Group (RTOG) 0522 study (**Figure 1a**) [45]. The CT data had a voxel resolution of 0.98 x 0.98 x 2.4 mm$^3$. Important structures such as the tumor, thyroid, lacrimal glands (both left and right), parotid glands (both left and right), and bone were contoured manually using Amira v.5.3.3 (**Figure 1b**). The labels were then converted to .stl files, which is a file format native to the stereolithography CAD software created by 3D systems and required for 3D printing. The .stl files were then modified for final assembly using Magics (Materialise NV, Leuven, Belgium), which is an .stl editing software package (**Figure 1c,d**). Modifications included the addition of flanges to seal the main volume, the incorporation of mounting points for the assembly of bones and other anatomical features, and ports for filling and sealing the tumor and gland voids.

The final phantom design consisted of nine custom parts, along with commercial hardware for assembly. This included seven 3D printed pieces, made with two different materials, and two laser-cut acrylic plates. The main chamber (skin) was printed using stereolithography (SL) out of Accura60, a proprietary photopolymer resin (3D Systems, Rock Hill, SC, USA), and then clear-coated for water resistance. The thyroid gland was fused to the skin and can be filled independently. The skull and spine were split into three separate parts and printed using SL out of Somos PerFORM, a proprietary photopolymer resin (DSM, Heerlen, Netherlands). The lacrimal glands were fused to the front piece of the skull and can be filled independently. Lastly, the parotid voids and tumor void were printed using SL out of Accura60 (3D Systems). The top and bottom of the chamber were formed with quarter inch acrylic plates that were laser-cut to their final size with a PLS6.75 Laser Cutting System (Universal Laser Systems, Scottsdale, AZ, USA). Commercial hardware included o-rings for sealing and nylon bolts for assembly of separate components. The final 3D printed head-and-neck anthropomorphic phantom is shown in **Figure 1e** and a sagittal slice of a CT image of the phantom is depicted in **Figure 1f**. **Table 1** shows the physical material properties taken from material datasheets of the Accura60 and Somos PerFORM comprising the phantom.

**Table 1**: Physical properties of the bulk materials

| Material | Tensile strength (MPa) | Elongation at break (%) | Tensile Modulus (MPa) | Flexural strength (MPa) | Water absorption (%) | Hardness (Shore D) |
|---|---|---|---|---|---|---|
| **Accura60** | 58-68 | 5-13 | 2690 - 3100 | 87-101 | N/A | 86 |
| **Somos PerFORM** | 68-80 | 1.1-1.2 | 9800 - 10500 | 120-146 | 0.1-0.2 | 93-94 |

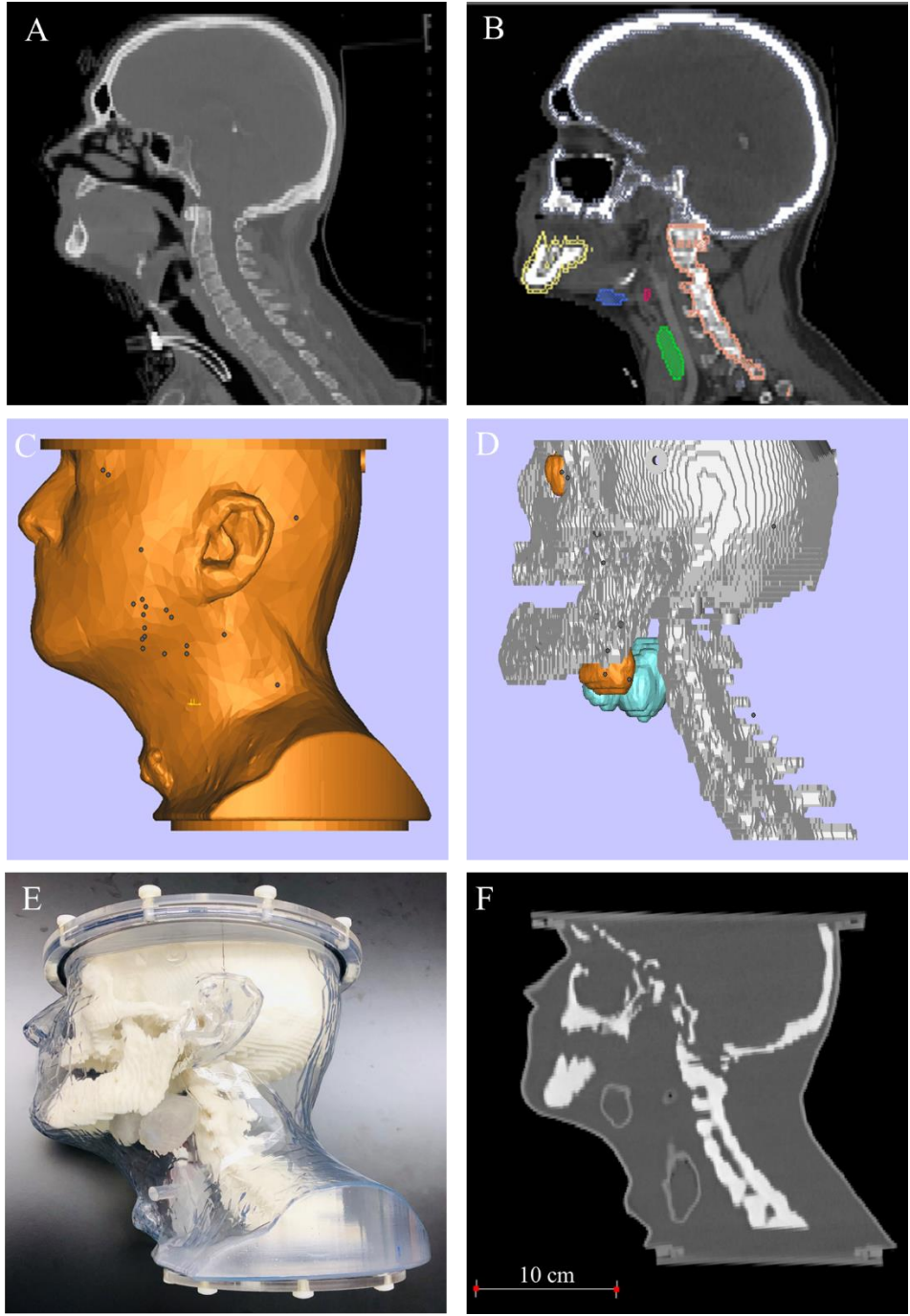

**Figure 1:** Representation of the H&N anthropomorphic phantom development process. (a) Original CT image obtained from a RTOG managed publicly available image repository and (b) relevant contoured structures. (c-d) 3D rendering of patient's skin, bony anatomy, and fillable organ anatomies. (e) Final 3D printed H&N anthropomorphic phantom. (f) CT image of H&N anthropomorphic phantom.

## 2.2 Phantom studies and SPECT/CT image acquisition

Three phantom studies were conducted, the parameters of which are summarized in **Table 2**. The first was conducted using a uniform 20 cm cylindrical phantom of volume 5.64 L to calculate the calibration factor to convert the reconstructed SPECT images to units of activity concentration. The second study was with a Jaszczak phantom containing 0.5-, 1-, 2-, 4-, 8-, and 16-ml hollow sphere inserts, with corresponding diameters of 9.9, 12.4, 15.4, 19.8, 24.8 and 31.3 mm, injected with a 9.9:1 insert to background activity concentration and was used to calculate spherical based recovery coefficients (RC) for PVC determination. The third study was with the anthropomorphic phantom containing a 22.8 ml tumor insert and injected with a 9.9:1 insert to background activity concentration. The resultant images were used to calculate an anthropomorphic based recovery coefficient and served as the basis for the image based RPT dosimetry calculation. SPECT/CT acquisitions were conducted using a GE Optima NM/CT 640 with a high energy general purpose (HEGP) collimator. All images were acquired over 360 degrees into 128x128 pixel matrices per angle. Body contouring was enabled. A photopeak energy window centered at 364 keV and 20% in width (327.6 – 400.4 keV) and a triple energy window scatter correction with two 20% windows adjacent (267.3 – 326.7 keV and 401.4 – 490.6 keV) to the photopeak window were used in the acquisition. A CT was acquired after the SPECT acquisition for CT-derived attenuation corrections (Hawkeye CT, 120 kVp, 20 mA, 487 mAs) and for accurate determination for the location of hot spheres and tumor insert.

Raw image data was reconstructed using the GE Xeleris 4.0 ordered subset expectation maximum (OSEM) algorithm with 10 iterations and 10 subsets with CT-based attenuation correction and both with and without GE's CDR modeling named 'Resolution Recovery' (without: non-RR, with: RR). No post-reconstruction filtering was applied. The reconstructed SPECT image matrices were 128 x 128 x 128 voxels with a voxel size of 4.42 x 4.42 x 4.42 mm$^3$ and the reconstructed CT images had a voxel size of 0.98 x 0.98 x 5.0 mm$^3$. High-resolution CT images of the anthropomorphic phantom were also acquired using a Siemens SOMATOM Definition Edge (120 kVp, 452 mA, 226 mAs) with a voxel size of 0.977x0.977x2.0 mm$^3$.

**Table 2**: Parameters used for SPECT/CT phantom studies

| Study | Initial $^{131}$I Activity Concentration (kBq/ml) | Insert to Background Ratio | Number of frames | Seconds per frame |
|---|---|---|---|---|
| **20 cm cylindrical phantom** | 105 | N/A | 120 | 20 |
| **Hot sphere Jaszczak** | 259 | 9.9:1 | 120 | 60 |
| **Anthropomorphic** | 257 | 9.9:1 | 120 | 60 |

### 2.3 Calculation of calibration factor and recovery coefficients

To convert the recorded SPECT image data into units of activity, the calibration factor, Q, was calculated according to Equation (1),

$$Q = \frac{R}{V \cdot C} \cdot e^{\lambda \Delta t} \quad (1)$$

where R is the mean count rate of a voxel in a VOI, V is the volume a voxel, C is the activity concentration at the time of syringe measurement, and the last term accounts for physical decay between source preparation time and scan time (Δt).

If $R_{voxel}$ is the count rate in a voxel, then the activity in the voxel, $A_{voxel}$, is given by Equation (2).

$$A_{voxel} = \frac{R_{voxel}}{Q} \quad (2)$$

To account for PVE, recovery coefficients were calculated according to Equation (3) where the measured activity concentration in the object VOI was determined by quantifying the mean value of each hot insert. Contours for the hot inserts were drawn on the CT and then the reconstructed SPECT was up-sampled using a Lanczos filter to the resolution of the CT and registered to the CT.

$$RC = \frac{\text{measured activity concentration in object VOI}}{\text{true activity concentration in object VOI}} \quad (3)$$

To correct the activity in a voxel, the RC is then used in Equation (4) where $R_{voxel}$ is the count rate in the voxel.

$$A_{voxel} = \frac{R_{voxel}}{Q \cdot RC} \quad (4)$$

The recovery coefficients for the Jaszczak spheres were then fit to the function given in Equation (5) where $R_{plateau}$, β, and γ are the curve fitted parameters [46].

$$RC_{fit} = R_{plateau} - \frac{R_{plateau}}{1 + \left(\frac{V}{\beta}\right)^{\gamma}} \quad (5)$$

**2.4 Monte Carlo dose calculation**

An in-house radiopharmaceutical dosimetry platform called RAPID was used to calculate the mean absorbed dose to the tumor structure in the anthropomorphic phantom [30]. RAPID utilizes nuclear medicine images to define the radionuclide activity in each voxel and the absorbed dose rate distribution is calculated on the CT which defines the material composition and mass density of the simulation geometry. The acquired SPECT/CT data from the anthropomorphic phantom study was used to calculate a voxel-level dose rate distribution in the phantom. The dose rate in each voxel was then integrated assuming simple exponential decay to produce a total absorbed dose distribution in the phantom.

Three scenarios were considered: i) ground truth using a simulated, idealized, SPECT dataset, ii) SPECT-based dosimetry without RCs, and iii) SPECT-based dosimetry with anthropomorphic phantom informed RCs. The ground truth scenario was run by creating an artificial activity distribution assuming a uniform concentration of 26.2 kBq/cc and 259 kBq/cc in the background and tumor insert respectively. The SPECT data was interpolated to the same resolution as the CT. The dose calculation grid was of dimensions 278 x 280 x 155 voxels with a 0.977x0.977x2.00 mm³ voxel size. The simulation was run with enough particles (8000 decays per voxel) to ensure Monte Carlo statistical uncertainty below 1.0% in the tumor region. Simulations were run on the UW Center for High Throughput Computing (CHTC).

**2.5 Uncertainty analysis**

Uncertainty analysis was performed according to the recommended EANM guidelines for determining error propagation in radiopharmaceutical dosimetry [47].

The fractional uncertainty of the volume determination is given by Equation (6) where a is the voxel size and D is the equivalent sphere diameter of the contoured structure.

$$\frac{u(v)}{v} = 3\left(\frac{\sqrt{\frac{a^2}{6}}}{D}\right) \quad (6)$$

The uncertainty of the calibration factor (calculated using Equation (1)) is given by Equation (7) and combines in quadrature two components: the assumed 5% error of the radionuclide calibrator activity measurement ($u(A_{cal})$), and the error of the count rate for the scans to determine the calibration factor $u(C_{ref})$, taken to be the standard deviation of the mean count rate for two timepoints.

$$\left[\frac{u(q)}{q}\right]^2 = \left[\frac{u(A_{cal})}{A_{cal}}\right]^2 + \left[\frac{u(C_{ref})}{C_{ref}}\right]^2 \quad (7)$$

The uncertainty of the expression used to fit the recovery coefficient curve (Equation (5)) is given by Equation (8). $\mathbf{g_c}$ is a matrix of dimensions 4 x 1 containing both the partial derivatives of first order of RC with respect to $R_{plateau}$, β, γ, and v, and $\mathbf{V_C}$ is a 4 x 4 matrix containing the covariance matrix from the least-squares fitting process and the uncertainty of the volumes, $u^2(v)$.

$$u^2(RC) = \mathbf{g_c^T V_c g_c} \tag{8}$$

The uncertainty in the measured mean counts C associated with the accuracy of the VOI definition is given by Equations (9) and (10),

$$\frac{u(C)}{C} = \frac{\varphi}{2RC} \frac{u(v)}{v} \tag{9}$$

$$\varphi = erf\left(\frac{2r^2}{\sigma\sqrt{2}}\right) - \frac{2\sigma}{r\sqrt{2\pi}}\left[1 - e^{-\frac{2r^2}{\sigma^2}}\right] \tag{10}$$

where r is D/2 (the equivalent sphere diameter of the structure) and σ is the standard deviation of the Gaussian point spread function derived from the system spatial resolution, assumed to be 3mm [48].

The cumulative activity uncertainties were calculated by adding the system sensitivity, recovery coefficient, and count rate errors in quadrature and subtracting out the covariance of the count rate and recovery coefficient on volume [34]. The uncertainty of the activity was estimated using Equation (11).

$$\left[\frac{u(\tilde{A})}{\tilde{A}}\right]^2 = \left[\frac{u(q)}{q}\right]^2 + \left[\frac{u(RC)}{RC}\right]^2 + \left[\frac{u(C)}{C}\right]^2 - \frac{\varphi}{RC^2 v}\frac{\partial RC}{\partial v}u^2(v) \tag{11}$$

### 3. Results
### 3.1 Phantom geometry

**Figure 2 a-c** shows photographs of the head and neck phantom and labels for the fillable and removable inserts. **Figure 2a** shows an image of the phantom assembled and filled with water. **Figure 2b** shows the anterior half of the skull insert with the 3D printed tumor volume and parotid glands attached to it by simple fasteners. **Figure 2c** shows a side profile of the phantom depicting the lacrimal glad (integrated into the skull) and the thyroid gland (integrated into the shell).

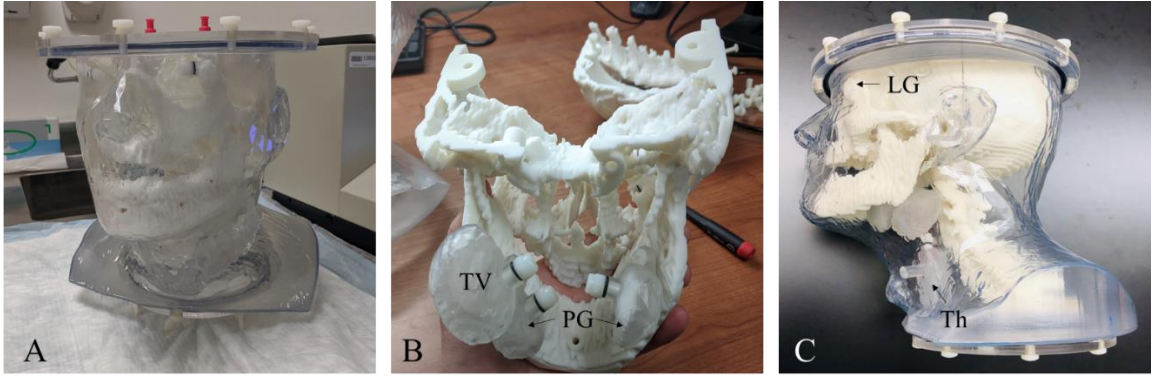

**Figure 2:** Photographs showing the (a) main chamber (skin), (b) lower jawbone assembly, and (c) compiled head and neck phantom. Labels for the fillable inserts including the tumor volume (TV), parotid gland (PG), lacrimal gland (LG), and thyroid (Th) are superimposed on the photographs.

### 3.2 Material Properties of Phantom

**Figure 3** shows axial, coronal, and sagittal CT slices of the phantom. **Figure 3a** shows an axial slice in which the jaw, spine, parotid glands, and tumor volume is present. **Figure 3b** shows a coronal slice of the phantom in which the skull and the C1-C5 vertebrae are present. **Figure 3c** depicts a sagittal view of the phantom in which the thyroid gland and the parotid gland are present.

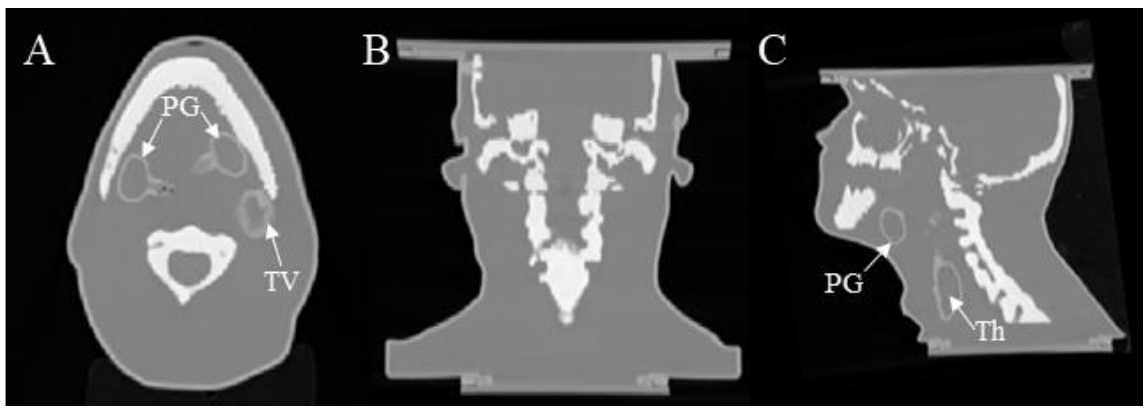

**Figure 3:** CT scan of phantom showing (a) axial, (b) coronal, (c) sagittal views. Labels for the fillable inserts including the tumor volume (TV), parotid gland (PG), and thyroid (Th) are superimposed on the photographs.

**Table 3** provides the radiological properties of the bulk materials. Densities were provided by the manufacturer. The measured HU corresponds to ROIs drawn on bulk portions of the material by the Siemens high-resolution CT scanner. The tabulated attenuation coefficients are calculated according to the bilinear relationships established by Brown et. al. for [131]I and PET [49] and Kabasakal et. al for [177]Lu [50].

**Table 3**: Radiological properties of the bulk materials

| Material | Density (g/cc) | Measured HU | Attenuation coefficient (cm$^{-1}$) | | |
|---|---|---|---|---|---|
| | | | $^{177}$Lu [50] | $^{131}$I (364 keV) | PET (511 keV) |
| **Accura60** | 1.21 | 292.48 +/- 8.57 | 0.1579 | 0.1063 | 0.1068 |
| **Somos PerFORM** | 1.61 | 876.36 +/- 14.77 | 0.2137 | 0.1390 | 0.1402 |
| **Water** | 1.00 | -4.05+/-7.95 | 0.1295 | 0.0896 | 0.0896 |

**Figure 4** depicts a portion of the CT calibration curve for the high-resolution CT scanner that the anthropomorphic phantom was scanned in. It also plots the two 3D printed materials alongside the curve in which the clinical calibration curve would slightly overpredict the electron density of the material.

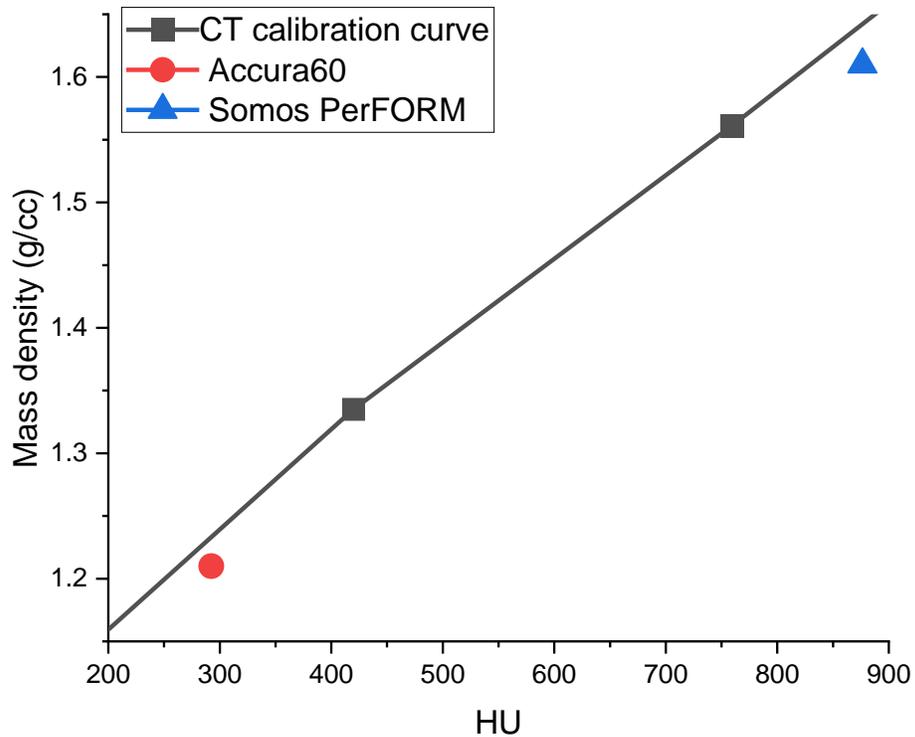

**Figure 4:** Portion of the CT calibration curve for the Siemens high resolution CT scanner, and the 3D printed materials overlaid

Table 4 shows a comparison between the measured volume of fillable chambers in the phantom and the volume of contours made on a high-resolution CT of the phantom. The maximum relative difference between the measured and contour is -10.07% for the right lacrimal gland which corresponds to an absolute difference of 0.15 ml. The largest absolute difference in volume was for the thyroid gland at 1.35ml. This is likely because it is the most irregularly, aspherical, shaped organ.

Table 4: Comparison of fillable volumes in anthropomorphic phantom (ml)

| Fillable void | Measured | DICOM contour | Absolute difference | Percent difference |
|---|---|---|---|---|
| Parotid$_L$ | 8.75 | 8.44 | -0.31 | -3.54 |
| Parotid$_R$ | 8.39 | 8.39 | 0.00 | 0.01 |
| Lacrimal$_L$ | 1.42 | 1.51 | 0.09 | 6.34 |
| Lacrimal$_R$ | 1.49 | 1.34 | -0.15 | -10.07 |
| Thyroid | 14.12 | 15.47 | 1.35 | 9.56 |
| GTV | 22.77 | 22.50 | -0.27 | -1.19 |

### 3.3 Phantom imaging studies

Using Equation (1), the calibration factor (Q) was calculated to be 104.2cps/MBq using RR and 21.4 cps/MBq for non-RR. **Figure 5** shows axial, coronal, and sagittal SPECT/CT slices of the anthropomorphic phantom after SPECT/CT acquisition. For the RR case, the average measured activity in the background and tumor compartments were 24.4 kBq/cc and 127.5 kBq/cc, respectively where the background compartment was a large ROI in the shoulder region containing only water. There was a -6.76% difference between the actual background (26.2 kBq/cc) and measured background. For the non-RR case, the average measured activity in the background and tumor compartments were 24.7 kBq/cc and 110.1 kBq/cc. The difference between the actual background and measured background was -5.81%.

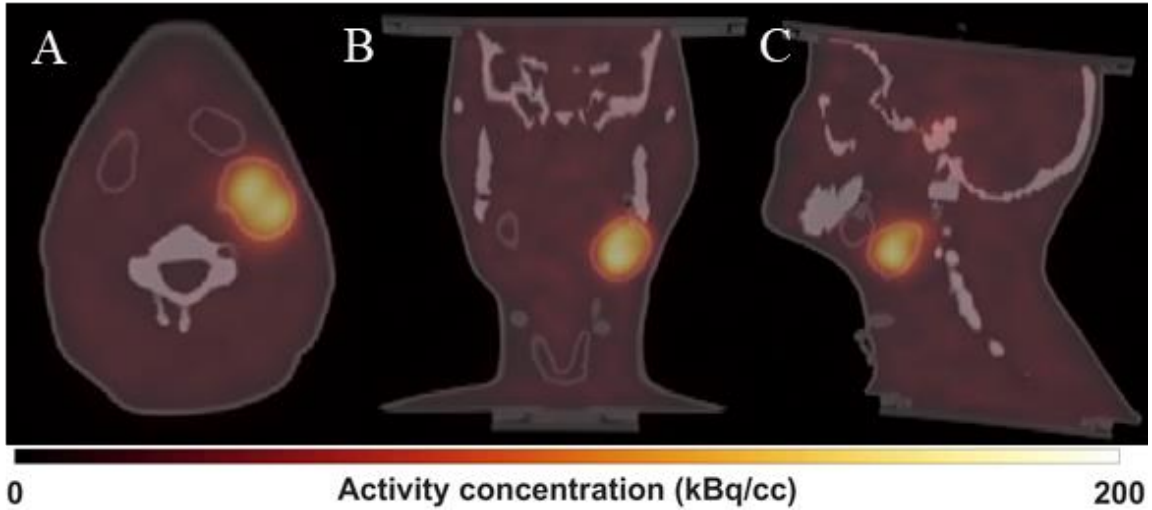

**Figure 5:** The reconstructed SPECT image and superimposed CT of the I-131 scan showing the (a) axial, (b) coronal, (c) sagittal views of the anthropomorphic phantom.

**Figure 6** depicts an axial SPECT/CT slice of the Jaszczak phantom after image acquisition and shows the difference in activity recovered with and without resolution recovery. For the RR case, the average measured activity in the background compartment and 16ml hot sphere compartments were 26.5 kBq/cc and 133.2 kBq/cc, respectively and there was a 2.35% difference between the actual background (25.9 kBq/cc) and measured background. For the non-RR case, the average measured activity in the background compartment and 16ml hot sphere compartments were 26.9 kBq/cc and 109.0 kBq/cc, respectively; and there was a 3.66% difference between the actual background (25.9 kBq/cc) and measured background.

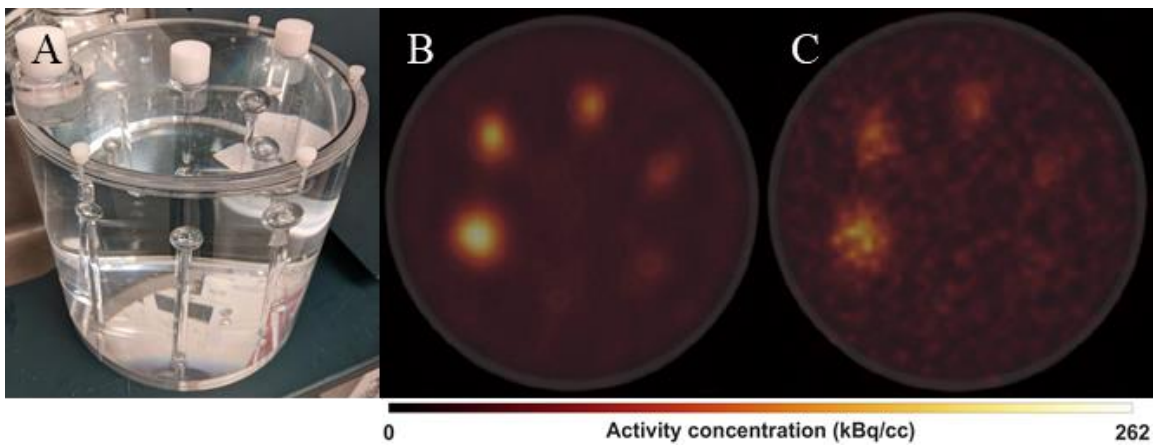

**Figure 6:** (a) Photograph of the Jaszczak phantom and (b) an axial slice of the reconstructed SPECT image with attenuation and scatter correction and superimposed CT of the I-131 scan with RR and (c) non-RR.

**Figure 7** shows the recovery coefficient for the Jaszczak phantom and 3D printed insert as a function of sphere insert size for both the non-RR and RR cases. As the size of the Jaszczak sphere increased, the recovery coefficients increased. Consistent with theory, the curve fit of the Jaszczak spheres in both cases with the non-RR and RR applied demonstrated that the spheres recovered more activity than the aspherical 3D printed volume [33]. The Jaszczak phantom results are fitted to the curvefit as in Equation (5). To calculate the recovery coefficient from the Jaszczak phantom for the anthropomorphic tumor volume, the fitted curve was extrapolated to the volume of the tumor volume. The extrapolated recovery coefficient was greater than that of the 3D printed insert for both the non-RR and RR cases (with RR: 0.536 vs. 0.493, non-RR: 0.445 vs. 0.426 respectively).

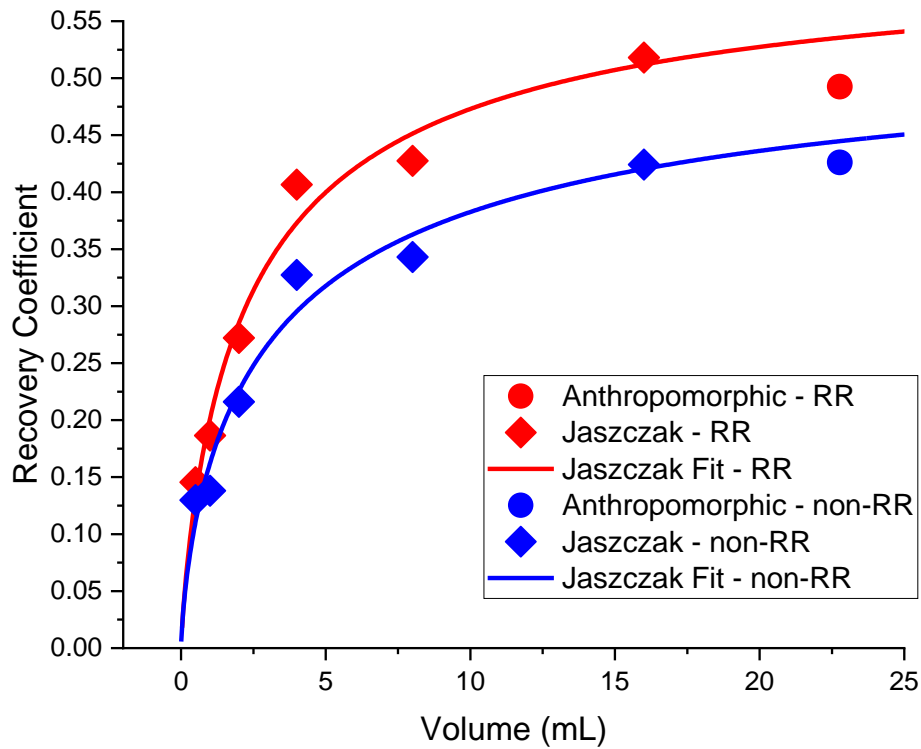

**Figure 7:** Recovery coefficients calculated for the Jaszczak phantom and 3D printed insert as a function of feature size for both RR and non-RR.

### 3.4 Monte Carlo dosimetry

**Figure 8** shows the difference in the calculated mean tumor dose in comparison to the ground truth scenario with an absorbed dose calculated to be 62.15 Gy. For the RR (non-RR) case, the mean dose to the tumor was underestimated by 46.8% (53.1%) before applying any RC, underestimated by 4.9% (0.7%) after applying a Jaszczak informed RC, and overestimated by 3.0% (3.0%) after applying the 3D printed insert RC.

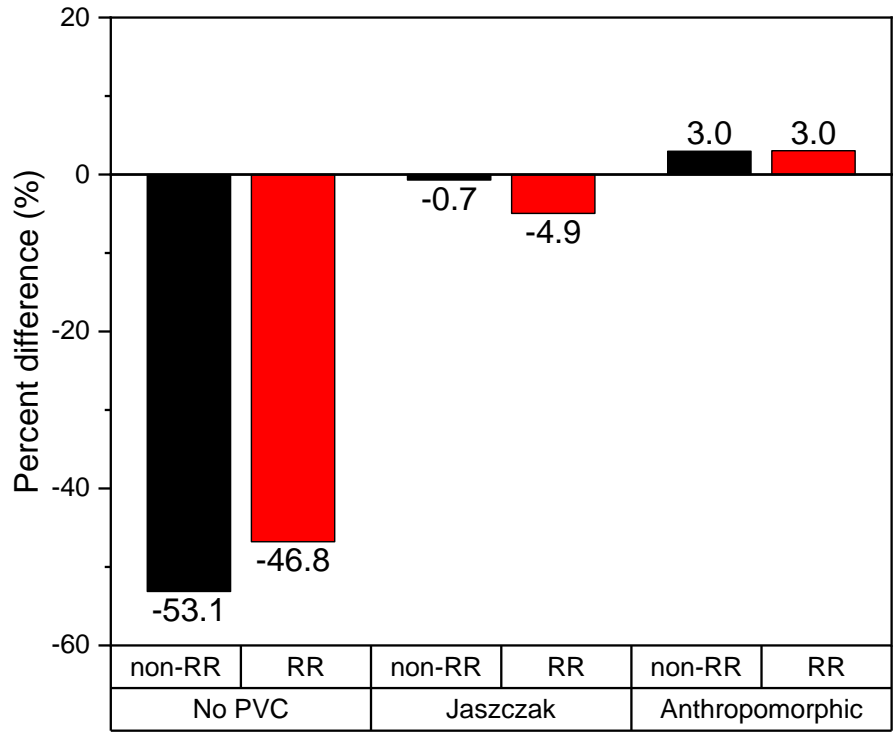

**Figure 8:** Percent difference in mean absorbed dose to the tumor volume compared to the ground truth (GT) scenario of 62.15 Gy for the scenarios considering RR and PVC.

**Figure 9** shows (a) axial, (b) coronal, and (c) sagittal slices of the absolute absorbed dose rate for the ground truth scenario and the respective (d-f) relative percent differences calculated between the ground truth and the SPECT-derived activity distributions for the resolution recovery case.

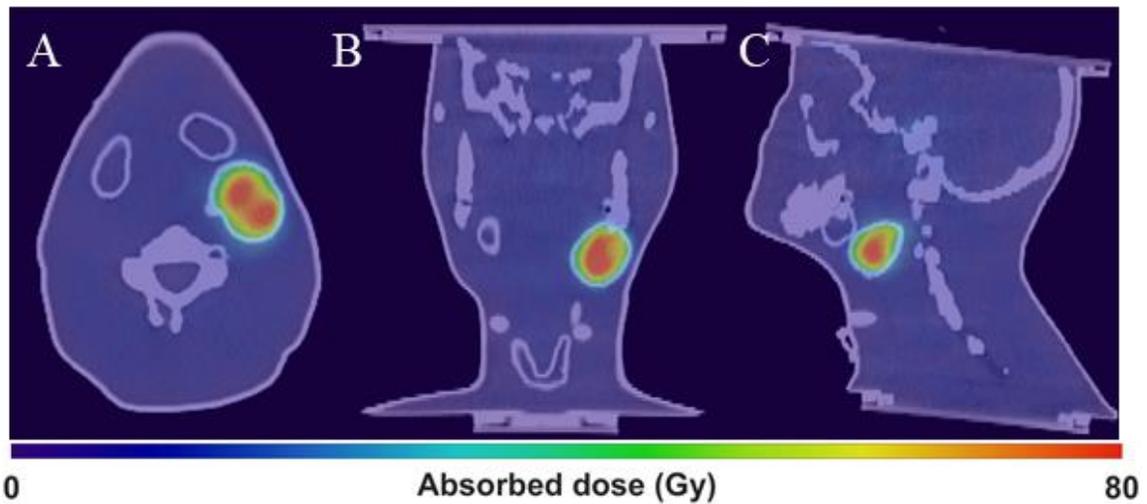

**Figure 9:** Absolute absorbed dose for the resolution recovery scenario: (a) axial, (b) coronal, and (c) sagittal views.

The calculated uncertainties for the RR case are given in
**Figure 10**. The largest uncertainties were the mean counts and RC which are dependent on the volume determination and the RC fitting function. The range of uncertainties for volume (17.3% to 54.7%), RC (22.32% to 95.71%), mean counts (13.2% to 106.4%), calibration factor (7.47%), and cumulative activity (25.6% to 113%) were calculated.

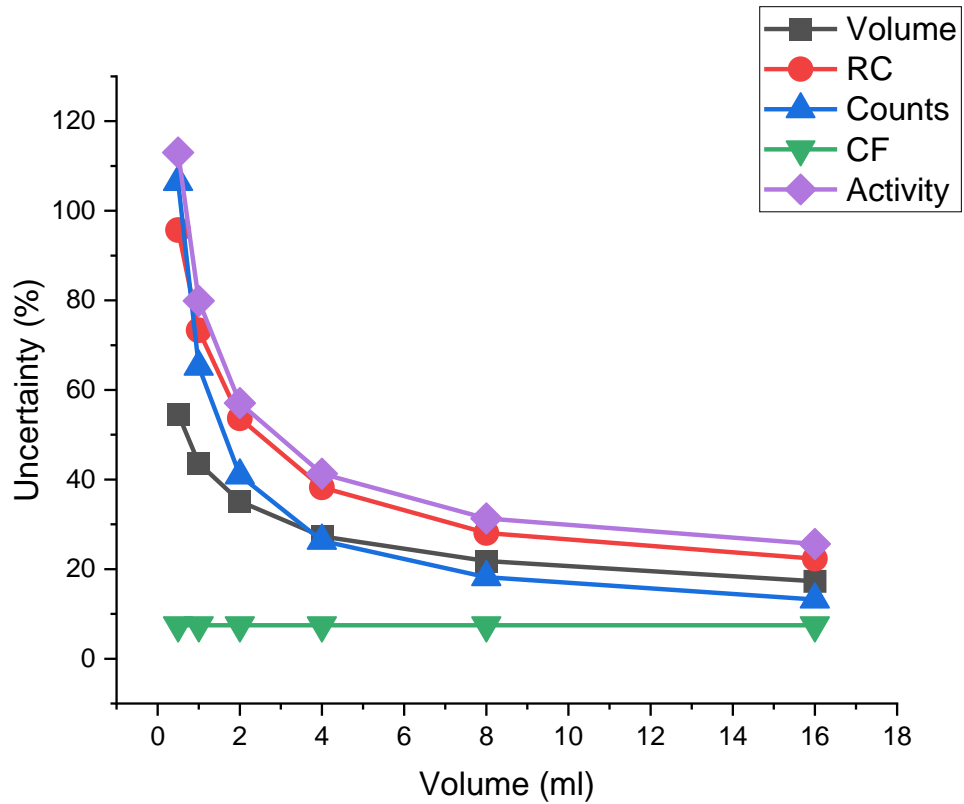

**Figure 10:** Uncertainties from Jaszczak phantom hot spheres study as a function of sphere volume.

## 4. Discussion

It was found that, in this study, the Jaszczak based PVC were sufficient to correct the activity distributions. While the Jaszczak-derived RC for the non-RR reconstruction was the most accurate scenario, the anthropomorphic phantom derived RC was most accurate for the RR reconstruction, within 3% of the ground truth. The tumor volume studied was quite spherical in shape and it is possible that for more aspherically shaped volumes, an anthropomorphic RC may be necessary. In the context of uncertainty analysis, the uncertainty in the activity determination was found to be around 25% for the 16 ml Jaszczak sphere, placing all the recovered activity dosimetric results within the calculated error and shows that the RCs determined from the anthropomorphic phantom were dosimetrically similar to those derived from the Jaszczak hot sphere phantom. The uncertainties were even higher (up to >100%) for smaller Jaszczak volumes. This is consistent with previous work in which VOI delineations have a substantial impact on the activity quantification and dose which are especially important for small structures [48].

Additionally, uncertainties have been reported to be up to 102% for small targets for $^{177}$Lu in human clinical trials, placing our results in line with other RPT uncertainties [51].

Because the calibration factor depends on the accuracy of scatter and attenuation correction, and they vary with phantom geometry[48], the shape of the phantom can directly affect the accuracy of quantifying activity in the background compartment. This may be the reason for the calibration factor slightly underpredicting the activity concentration in the background compartment (an ROI in the shoulder containing only water) of the anthropomorphic phantom. These types of shape-dependent calibration factors have been previously reported [37,39]. The calibration factor derived from the homogeneous Jaszczak phantom study was used to keep the anthropomorphic phantom study in line with the imaging parameters and recommended guidelines for upcoming clinical trial patient studies [34].

In addition to the difference in shape of the phantom's background region, another important distinction of the anthropomorphic phantom is the inclusion of 3D printed bone-mimicking material which more resembles the density and corresponding HU of bones in the head and neck to afford a more realistic dosimetric scenario. The bone-type material was measured to have a bulk HU value of 876.36 and density of 1.61 g/cc which is representative of cortical bone, oftentimes found in the skull. Uncertainties and differences of 3D printed materials have been demonstrated in previous work by Craft et.al [52]. In that work, slight differences in density and composition of printed materials resulted in dosimetric differences for clinical photon and electron beams. In the context of $^{131}$I RPT dosimetry, not only are dose computations affected because 3D printed materials may not lie on the CT calibration curve (see **Figure 4**), but CT-based attenuation corrections for image reconstruction may be affected as well. These material property uncertainties could affect the dosimetry and the further characterizing of 3D printed materials should be carefully considered.

RR, the CDR compensation technique included in GE's reconstruction software, Xeleris 4.0 [53,54] was investigated because, in theory, high energy photon emissions of $^{131}$I requires modeling of the high energy collimators that inherently reduce spatial resolution due to the collimator geometry. We found that there was only a small difference in the dosimetric results between non-RR and RR; however, utilizing RR increased the spatial accuracy of the activity quantification and has been recommended to be used[34]. Additionally, the spatial distribution of activity was less clear and difficult to discern the presence of activity in smaller structures (see **Figure 6c**) thus limiting its applicability. The CDR model in Xeleris only models the intrinsic and geometric components of the collimator and neglects the collimator scatter and penetration, which can lead to significant higher count recovery for $^{131}$I [34]. In addition to using the manufacturer's RR, Monte

Carlo CDR modeling has been studied in previous work to model the CDR explicitly to account for septal penetration and scatter [55,56] and should be considered to improve the accuracy of the image reconstruction.

Large dose gradients were present in the RR dosimetry. This is consistent with both spill out due to imaging system limitations and Gibbs ringing artifacts that have been documented when using CDR modeling[57,58,35]. As depicted in relative difference maps in **Figure 11**, there are stark regions of red, indicating that the SPECT image has much higher dose than the corresponding ground truth. This is attributable to those certain structures like bone material and 3D printed material were assumed to be impermeable to water and thus activity was not assigned to them in the ground truth scenario.

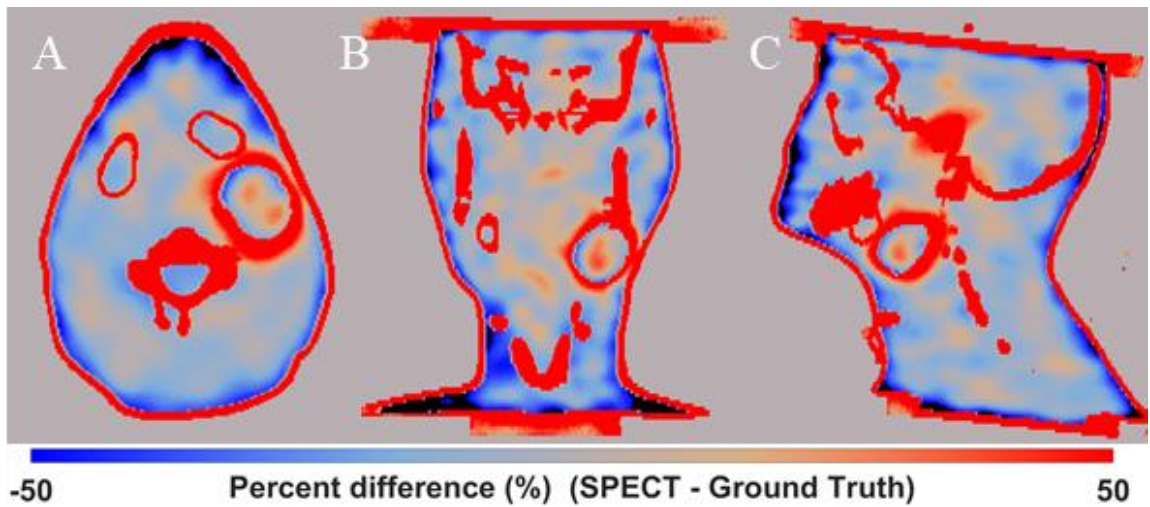

**Figure 11:** Relative dose differences between the ground truth scenario and SPECT-derived activity distribution with resolution recovery for (a) axial, (b) coronal, and (c) sagittal views.

## 5. Conclusions

The 3D printed anthropomorphic phantom was created to more accurately represent the geometry and material composition of the head and neck region in comparison to typical non-anthropomorphic Jaszczak-type nuclear medicine phantoms. This phantom was used to evaluate the accuracy of a Monte Carlo, image based, RPT dosimetry workflow in comparison to a ground truth scenario. We investigated the determination of RCs from different phantoms, the impact of CDR modeling, and assessed the uncertainty in the activity determination. After correcting the reconstructed SPECT images using volume-based RCs, the dosimetry workflow was determined to be accurate within the calculated uncertainty. The Jaszczak based PVC were sufficient to correct the activity distributions in this study; however, aspherical volumes may warrant the utilization of anthropomorphic phantom PVC. Further work should characterize the 3D printed materials rigorously and include Monte Carlo based reconstructions which include more accurate CDR modeling.

**Acknowledgements and Financial Disclosure**

We would like to thank the UW CHTC for the use of their cluster and their computational support. This project was partially supported by the Morgridge Institute for Research, the Specialized Program of Research Excellence (SPORE) program, through the NIH National Institute for Dental and Craniofacial Research (NIDCR) and National Cancer Institute (NCI), grant P50DE026787, the National Cancer Institute (NCI) Research Project--Cooperative Agreements U01CA233102-01, and the NIH National Cancer Institute (NCI), grant P01CA250972. The content is solely the responsibility of the authors and does not necessarily represent the official views of the NIH.

**Disclaimer**

BLC is a co-founder of Phantech LLC, a manufacturer of imaging phantoms. BB and JG are co-founders of Voximetry, Inc., a nuclear medicine dosimetry company in Madison, WI.

26